\documentclass[sigconf, pbalance]{acmart}

\usepackage[capitalize, noabbrev]{cleveref}
\usepackage{upgreek}
\usepackage{todonotes}
\usepackage{listings}
\usepackage{subcaption}

\copyrightyear{2026}
\acmYear{2026}
\setcopyright{cc}
\setcctype{by}
\acmConference[L@S '26]{Proceedings of the Thirteenth ACM Conference on Learning @ Scale}{June 29-July 03, 2026}{Seoul, Republic of Korea}
\acmBooktitle{Proceedings of the Thirteenth ACM Conference on Learning @ Scale (L@S '26), June 29-July 03, 2026, Seoul, Republic of Korea}
\acmDOI{10.1145/3774398.3811592}
\acmISBN{979-8-4007-2293-6/2026/06}

\begin{document}

\title{$\upmu$Ed API: Towards a Shared API for Education Microservices}


\author{Maximilian S\"olch}
\authornote{These authors contributed equally to this work.}
\email{maximilian.soelch@tum.de}
\orcid{0009-0004-1509-7842}
\affiliation{%
  \institution{Technical University of Munich}
  \city{Munich}
  \country{Germany}
}

\author{Alexandra Neagu}
\authornotemark[1]
\email{alexandra.neagu20@imperial.ac.uk}
\orcid{0009-0004-0840-3776}
\affiliation{%
  \institution{Imperial College London}
  \city{London}
  \country{UK}
}

\author{Marcus Messer}
\authornotemark[1]
\email{m.messer@imperial.ac.uk}
\orcid{0000-0001-5915-9153}
\affiliation{%
  \institution{Imperial College London}
  \city{London}
  \country{UK}
}

\author{Peter B. Johnson}
\authornotemark[1]
\email{peter.johnson@imperial.ac.uk}
\orcid{0000-0001-7841-691X}
\affiliation{%
  \institution{Imperial College London}
  \city{London}
  \country{UK}
}

\author{Gerd Kortemeyer}
\email{kgerd@ethz.ch}
\orcid{0000-0001-6643-9428}
\affiliation{%
  \institution{ETH Z\"urich}
  \city{Z\"urich}
  \country{Switzerland}
}

\author{Samuel S. H. Ng}
\email{samuel.ng@ntu.edu.sg}
\orcid{0009-0006-3594-8178}
\affiliation{%
  \institution{Nanyang Technological University}
  \city{Singapore}
  \country{Singapore}
}

\author{Fun Siong Lim}
\email{lim_fun_siong@ntu.edu.sg}
\orcid{0000-0001-8887-6047}
\affiliation{%
  \institution{Nanyang Technological University}
  \city{Singapore}
  \country{Singapore}
}

\author{Stephan Krusche}
\email{krusche@tum.de}
\orcid{0000-0002-4552-644X}
\affiliation{%
  \institution{Technical University of Munich}
  \city{Munich}
  \country{Germany}
}

\renewcommand{\shortauthors}{S\"olch et al.}











\begin{abstract}

Learning at scale often requires domain-specific automation such as assessment and feedback. An organization locked in to a general learning platform without these specialist automations limits its pedagogical offering. An ecosystem of interoperable, platform-agnostic microservices for domain-specific automation would solve this problem. To develop an effective ecosystem, a standard interface (API) for education microservices is required.

We propose an initial specification for a standard, platform-independent API for educational microservices, $\upmu$Ed. The API integrates functionality from existing systems in use at four institutions, which are adopting the new API. The API is initially specified for automation of feedback, assessment, and educational chatbots, with further service types planned.
 
The API specification provided here enables the development of an ecosystem of education microservices that will facilitate automation in more domains, to more users, providing a richer learning experience in a wide range of disciplines. 

\end{abstract}

\ccsdesc[500]{General and reference~Computing standards, RFCs and guidelines}
\ccsdesc[500]{Applied computing~Education}

\keywords{EdTech, Education Technology, API, Automated Assessment, Automated Feedback, Educational Chatbots}

\maketitle

\section{Introduction}
Integrating an education software platform into organizational workflows is a complex and multi-staged process \cite{granic2022educational}. After integration, teachers and institutions can become dependent, or `locked-in' to a specific platform \cite{Pangrazio02012023}\cite[p.~140]{thomas2026critical}. 

Lock-in is particularly problematic when \textit{domain-specific} technologies are required. For example, automating feedback on a literature essay, or on a mathematical proof, requires domain expertise. Learning platforms are developed for general application, and cannot conceivably meet all the diverse specialist needs across academia. Lock-in therefore prevents pedagogical autonomy \cite{kerssens2022governed} and creates barriers to technical innovation \cite{baldwin2000design,Pangrazio02012023}.

A common approach to manage lock-in is for platforms to connect to external software applications, plugins, or services. External software can range in complexity from comprehensive applications, such as \textit{Turnitin}~\cite{batane2010turning} or \textit{M\"obius}~\cite{mobius}, to separate granular services, known as microservices \cite{lewis2014definition,Dragoni2017}. Examples of microservices include generating domain-specific feedback on tasks for students~\cite{lundengaard2024automated,liu2016automated}, or chat interactions for students or teachers~\cite{lai2025NALA, VACALOPOULOU2024AI4}. This paper focuses on microservices due to their applicability to the automation of domain-specific processes. 

The granularity of microservices lowers the barrier to entry, enabling subject experts to contribute specialist knowledge to individual services \cite{baldwin2000design}. An ecosystem of such microservices, connecting to learning platforms, would produce richer automation and greater scalability across platforms \cite{johnson2025microservices}.

To realize such an ecosystem requires a standard interface. In this paper, we specify the $\upmu$Ed API for education microservices, collaboratively developed and now being adopted on four existing systems and open for wider adoption.

\section{Related Work}\label{sec:related_work}

\subsection{Education Standards}
\citet{Bakhouyi2017} reviewed standardization and interoperability in e-learning systems, including three interfaces whose scope goes beyond learning content and includes protocols and interactions. LTI standardizes protocols between platforms and external learning applications \cite{lti}; SCORM consists of an overview, content aggregation model, runtime environment and sequencing and navigation standards \cite{scorm}; and xAPI, the newer version of SCORM, focuses on tracking learning experiences and activities \cite{xapi}. 

The existing standards emphasize the structure, or schema, of the data moving between systems, but do not prescribe an ontology, or the conceptual meaning of the data. This separation, which is appropriate for the heterogeneous and context-depend values in the education sector, should be retained in any new API designs.

Another commonality of existing standards is to integrate deeply into organizational workflows; for example, external tools can modify data within a platform. There are no standards for microservices with a simple request/response schema and minimal integration into client workflows --- such an API would require a new definition.


\subsection{Microservices}

Microservices are a general pattern that favors more granular, specialist services \cite{lewis2014definition, Dragoni2017}. A focus on granular tasks can lower barriers to entry, diversify sources of innovation, and lead to an ecosystem of services that enrich the offerings available and enhance their scale of deployment \cite{fullan2014rich,reigeluth2013reinventing,Otto2022,network2019scaling}. 

With higher-resourced institutions more likely to develop such services \cite{Buckingham2023}, microservices promote more equitable access to technology in education \cite{Gottschalk2023}. The microservice pattern can also address lock-in \cite{Pangrazio02012023} and help with transparency and accountability challenges \cite{Holmes2022, Klimova2023, FU2024100306}. The recent increase in automation of pedagogical judgments, as opposed to deterministic administrative decisions \cite{Selwyn2023}, especially using generative AI \cite{Zhang2024}, increases the importance of decentralized microservices.

In the absence of an existing standard for microservices in education, and with the desire to interoperably employ each others' services, four institutions represented in this paper collaborated to develop a new API standard.

\subsection{API Design}

A standard API aims for high adoption. The level of specification is a balance. Incorporating the needs of many stakeholders may lead to over-specification and low adoption; however prioritizing increased adoption may lead to under-specifying, which is counter to the goal of interoperable services \cite{Bloch2006, Stylos2007}. 

Clear and consistent terminology is also important to API design, even at the expense of preferred local vocabulary \cite{Bloch2006,Stylos2007}; existing education interfaces, including LTI, SCORM, and xAPI, follow this pattern. For example, the LTI term ``course'' could in other local contexts be ``module'' or ``class'', but it is more important to be consistent than to be locally representative. 

Deployment brings other API considerations, such as ensuring client identification and rotating keys for security, discovering capabilities, and avoiding over-fetching \cite{stocker2018interface}.

The conclusion from related work is that a specialist API for education microservices is required, that it should prioritize consistent terminology, manage the tension of over- vs under-specifying, and consider how the API will be utilized in deployment.

\section{Method}\label{sec:method}

The $\upmu$Ed API was collaboratively developed by four institutions based on existing systems and the desire for interoperability. The aim was to define an API for a small set of core capabilities that capture common needs in modern learning platforms, and allow these capabilities to be provided by specialized (external) services.

\subsection{Design Principles}
Four design principles guided the development of the $\upmu$Ed API:

\textbf{Interoperability}: The API is a stable interface between learning platforms and specialized services, enabling reuse and standardization of components developed at different institutions. By comparing existing education APIs across multiple institutions, we abstracted a common terminology that supports multi-institution solutions \anon{\cite{johnson2025lambdafeedback,krusche2018artemis,kortemeyer2024ethel,lai2025NALA}}, provides clear and consistent terminology \cite{Bloch2006, Stylos2007} and minimizes the effort to adopt the API.

\textbf{Provider-agnostic}: 
The API avoids binding interactions to specific providers or technologies. Optional configuration mechanisms allow experimentation without compromising interoperability.

\textbf{Partial adoption}: Service providers may implement only a subset of the defined capabilities and explicitly declare their supported features. Partial adoption of the API lowers the barrier to entry and enables incremental adoption across institutions.

\textbf{Pedagogical flexibility}: The API facilitates common schemas or data structures, without prejudice to ontology or pedagogical application. Mandatory inputs are minimized to provide flexibility. Core requests require only essential information, such as submission or a sequence of chat messages, while richer educational context, such as learning objectives, assessment criteria, or user information and preferences, remains optional.
The flexibility allows the same interface to be applied in diverse instructional settings.

\subsection{Cross-Institutional Design Process}
Rather than starting from a purely theoretical model, the $\upmu$Ed API design was derived by comparing, abstracting, and aligning existing e-learning systems. Four institutions in four countries collaborated: Imperial College London with \anon{Lambda Feedback \cite{johnson2025lambdafeedback}}; Technical University of Munich with \anon{Artemis \cite{krusche2018artemis}} and \anon{Athena \cite{solch:2025:DirectAutomatedFeedback,solch:2026:ScalingAssessmentStudent}}; ETH Z\"urich with \anon{Ethel \cite{kortemeyer2024ethel}}; and Nanyang Technological University with \anon{NALA \cite{lai2025NALA}}.

The initial API draft was jointly developed by two institutions (\anon{Technical University of Munich and Imperial College London}) based on their existing automated feedback, assessment and chat services, with input from a total of five instructors, researchers and developers.
As a first step, the teams compared their existing, independently developed APIs to identify overlapping concepts, differences in terminology, and institution-specific assumptions.
This comparison revealed a shared core of functionality, alongside variations in required inputs and interface structure.

In a second step, a total of four instructors, researchers and developers from two further institutions (\anon{ETH Z\"urich and Nanyang Technological University}) reviewed the draft specification and contributed API descriptions from their systems.
The second step validated that the emerging $\upmu$Ed API could accommodate other use cases without requiring substantial redesign of established services, and provided valuable modifications for further generalization.
The cross-institutional review informed refinements of the data model and the definition of supported capabilities.
Conflicting ideas and terms were resolved through multiple rounds of discussion and refinement until our initial proposed API was complete. 


\section{Initial Proposed API}\label{sec:initial_api}
\begin{figure}[htbp]
  \centering
  \includegraphics[width=\linewidth]{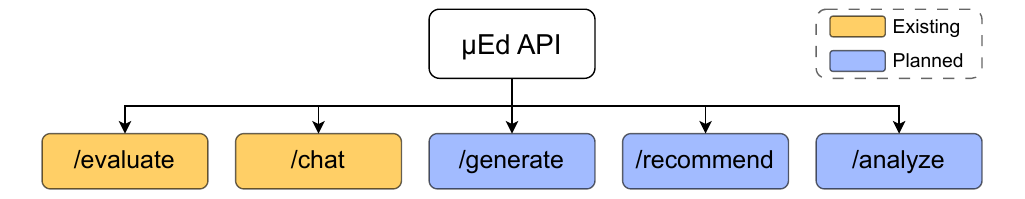}
  \caption{High-level overview of the $\upmu$Ed API, illustrating the five core capability domains (\texttt{/evaluate}, \texttt{/chat}, \texttt{/generate}, \texttt{/recommend}, \texttt{/analyze}) and their current development status.}
  \label{fig:med-api-overview}
\end{figure}

The collaborative approach produced $\upmu$Ed API v0.1, an interface for platform-independent education microservices. \cref{fig:med-api-overview} gives a high-level overview of the API,  structured around five core capabilities, each with a dedicated path. The first two are defined in the API: (1) automated evaluation of learner artifacts (\texttt{/evaluate}), (2) educational dialogue (\texttt{/chat}). The next three are planned capabilities that require defining: (3) educational content generation (\texttt{/generate}), (4) learning-oriented recommendation (\texttt{/recommend}), and (5) learning analytics (\texttt{/analyze}).

The specification is defined using the OpenAPI 3.1 \cite{openapi} standard, enabling both human-readable documentation and machine-processable integration. Example payloads are illustrated in \cref{tab:med-api-operations1} and \cref{tab:med-api-operations2}, and the full specification is provided as supplementary material~\cite{solch2026uedapi}. 

\vspace{-2mm}

\lstdefinelanguage{yaml}{
  keywords={true, false, null, yes, no},
  keywordstyle=\color{blue}\bfseries,
  sensitive=false,
  comment=[l]{\#},
  commentstyle=\color{gray}\itshape,
  morestring=[b]',
  morestring=[b]",
}

\lstset{
  language=yaml,
  basicstyle=\ttfamily\footnotesize,
  breaklines=true,          
  breakatwhitespace=false,  
  postbreak=\mbox{\textcolor{red}{$\hookrightarrow$}\space}, 
  columns=fullflexible,
  keepspaces=true,
  showstringspaces=false,
  rulecolor=\color{black},
  tabsize=2,
  numberstyle=\tiny\color{gray},
  numbersep=5pt,
}

\begin{table}[b]
\centering
\caption{Example payloads for \texttt{/evaluate}. For the full API specification, see~\cite{solch2026uedapi}.}
\label{tab:med-api-operations1}
\footnotesize
\begin{tabular}{p{0.5\linewidth} p{0.5\linewidth}}
\toprule
\textbf{Example Request}&
\textbf{Example Response} \\
\midrule
\vspace{-1em}
\begin{lstlisting}[language=yaml]
task:
  title: Explain polymorphism
  content:
    text: Define with one example.
  referenceSolution:    
    text: Polymorphism allows ...
  context:
    constraints: Answer in 3-6 sentences.
submission:
  content:
    text: Polymorphism means ...
criteria:
  - name: Correctness    
    maxPoints: 10
  - name: Clarity    
    maxPoints: 5
\end{lstlisting}
&
\vspace{-1em}
\begin{lstlisting}[language=yaml]
- feedbackId: fb-1
  title: Clarify your definition
  message: Explanation is generally correct. It would help to ...
  suggestedAction: Add one concrete example ...
  awardedPoints: 2.5
  criterion:
    name: Correctness
      
- feedbackId: fb-2
  title: Overall structure
  message: Your answer is easy to follow.
  criterion:
    name: Clarity
\end{lstlisting}
\\



\bottomrule
\end{tabular}
\end{table}

\begin{table}[b]
\centering
\caption{Example payloads for the \texttt{/chat}. For the full API specification, see~\cite{solch2026uedapi}.}
\label{tab:med-api-operations2}
\footnotesize
\begin{tabular}{p{0.55\linewidth} p{0.45\linewidth}}
\toprule
\textbf{Example Request}&
\textbf{Example Response} \\
\midrule

\vspace{-1em}
\begin{lstlisting}[language=yaml]
user:
  type: LEARNER
  detailPreference: MEDIUM 
messages: 
  - role: USER
    content: I do not know how to solve this.
context:
  task:
    title: Explain polymorphism
    content:
      text: Define with one example.
configuration:
  type: Java Assistant
  llm:
    model: gpt-5.2    
\end{lstlisting}
&
\vspace{-1em}
\begin{lstlisting}[language=yaml]
response:
  role: ASSISTANT
  content: Here's a hint for your answer: Focus on how polymorphism ...
metadata:
  responseTimeMs: 1800
  createdAt: 2026-02-16T10:30:00Z
  llm:
    model: gpt-5.2
\end{lstlisting} 
\\
\bottomrule
\end{tabular}
\end{table}

\subsection{Overview of Core Capabilities}
\subsubsection{\textbf{Evaluate (\texttt{/evaluate}):}}
A standardized interface for both \emph{formative feedback}, aimed at guiding learning, and \emph{summative assessment}, including the assignment of points or grades, on learner submissions.
The API does not enforce pedagogical approaches; service providers may implement only formative feedback, only summative assessment, or both, and explicitly communicate their supported features via capability metadata.

Conceptually, an \texttt{evaluate} request consists of a submission and optional contextual information, such as the task description, learning objectives, and assessment criteria.
This design allows services to operate with minimal input while enabling richer, more pedagogically grounded feedback when additional context is available.

A central design decision is the explicit support for different evaluation modes. In addition to complete evaluation, the API supports preliminary feedback, allowing the same interface to be used for rapid, lightweight feedback. For example, a preview of how handwritten submissions will be interpreted. Incorporating both functions into a single endpoint enables them to share functionality.


\subsubsection{\textbf{Chat (\texttt{/chat})}:}
Multi-turn conversations grounded in an educational context, such as task details, a submission, user preferences, course materials, or study goals. A \texttt{chat} interaction comprises a sequence of \textit{messages}, with an optional educational \textit{context} to adapt responses.
An optional conversation ID also enables stateful chat microservices containing their own memory of the user.

The current implementation is based on GenAI APIs, such as the OpenAI API \cite{openai_api} but intentionally avoids setting model- or technology-specific assumptions. For instance, the optional \textit{configuration} of the microservice contains a proposed structure for LLM-based chats, but is open to other technologies.
The API is not restricted to specific pedagogical interaction patterns or user types. For example, students asking for help, or teachers developing materials or querying policies, are equally permitted uses. 
This flexibility enables experimentation with different forms of conversational support while maintaining a shared interface across platforms.



\subsubsection{\textbf{Shared Aspects:}}
All core capabilities include optional authentication headers to enable user-specific access to services, an optional \textit{credentials} field to support sharing time-based keys provided by proxy services \cite{stocker2018interface}, and an optional \textit{dataPolicy} object to govern how services handle users' data, including legal basis and retention policies.

To avoid tight coupling between clients and service implementations, the API introduces a lightweight \texttt{/health} mechanism to discover supported features at runtime. The \texttt{evaluate} capability can declare which artifact types, languages, and evaluation modes are supported; the \texttt{chat} capability can declare supported features such as adaptation to user preferences or streaming responses. Health allows platforms to dynamically adapt their behavior and to integrate heterogeneous services without manual configuration.

\subsubsection{\textbf{Planned Capabilities (\texttt{/generate}, \texttt{/recommend}, \texttt{/analyze}):}}
We anticipate the addition of the capabilities to \texttt{generate} educational content, to \texttt{recommend} actions to a learner, and to \texttt{analyze} educational data. These capabilities are not part of the API specification released in v0.1, as we require more practical experience with these aspects. We encourage developers of these services or others to contribute API definitions for inclusion in future versions.

\newcommand{\muEd}{$\upmu$Ed}
\section{Discussion}\label{sec:discussion}
We have defined the $\upmu$Ed API for education microservices, now being adopted by our four institutions, achieving the initial goal of our work. We discuss our reflections on the design process, the identity of the API, its limitations, and future plans.

\subsection{Balance in the API Design}

To achieve an API with consistent terminology suitable for the four institutions required a constructive dialogue. An example is the decision to group formative feedback and summative assessment within the same \texttt{/evaluate} capability. These two processes \emph{can} be separate, but they are often intrinsically linked, so we combined them while maintaining the option to use only one or the other.

The API specification was balanced by including what was already required by the original systems, but not inventing details where it was not (yet) required. This approach recognized the level of maturity of the technology, prioritizing incremental levels of interoperability rather than seeking to fully specify. The current specification is operational but initial and limited. 

\subsection{Distinguishing From Other APIs}

The $\upmu$Ed API has a fundamentally lower level of integration into client workflows than comprehensive APIs such as LTI and SCORM. Microservices avoid, for example, the management of institutions, courses, and users. They do not manipulate data within the client platform. If those levels of integration are required then the scope of the external tool goes beyond a conventional `microservice’, and it is more appropriate to use existing interfaces such as LTI or SCORM. This discussion helps show the limits of microservices and how they have a particular role to play, rather than being a solution to all EdTech problems. The value of microservices is their decentralized origins and low barrier to entry, which are ideal for capturing the breadth of domain-specific knowledge required to improve automations that are sensitive to academic expertise and pedagogical norms. 



\subsection{Prospects for Adoption}

The vision of the API is that it becomes sector-wide, facilitating an ecosystem of microservices that bring domain-specific automations to a wider range of learners, enriching their learning, improving equity, and helping with transparency and accountability. 

The measures of success of the API in achieving these goals in future will be: (1) wider adoption by different platforms and services; (2) the number of calls from numerous services using the API; (3) genuine substitutions of services, as evidence of innovation and pedagogical freedom despite platform lock-in; (4) improving learner experience. The current API specification is a work in progress that provides the first step towards these ambitions. 



\subsection{Limitations and Future Plans}
The API is currently limited to two capabilities, (\texttt{/evaluate} and \texttt{/chat}), but is an extensible framework as illustrated by the anticipated \texttt{/generate}, \texttt{/recommend}, and \texttt{/analyze} capabilities. All capabilities share a common emphasis on 
optionality of inputs, and explicit communication of supported features.

The API meets the needs of four different e-learning platforms each from different institutions and countries. All four are adopting the API specification. The breadth of inputs ensures a more general approach than a single institution, but the bias towards the needs of these research-intensive, STEM institutions remains. 

Wider adoption will require input from the broader community, including  domains outside STEM and varying levels of education. Broadening the application may uncover new conflicts of terminology and purpose. We invite contributions to refine and improve the API in the same spirit articulated here, i.e.\ abstracting existing workflows; defining clear and consistent terminology; seeking balanced specifications levels; and considering practical deployment details. 

\section{Conclusion}

An ecosystem of platform-agnostic, interoperable education microservices can lower barriers to entry for domain-specific automations. Microservices facilitate more innovation, wider deployment, and improved transparency and equity. The result will be richer learning experiences scalable to more users in the sector. A key enabler to this vision is a standard API.

We presented the $\upmu$Ed API for education microservices, defined by combining the workflows of four institutions, which are now adopting the API. We invite developers to adopt the API, suggest extensions and improvements, and extend the interoperability across more platforms and services. The full API is defined in \cite{solch2026uedapi}.

\begin{acks}
In accordance with ACM author policy, we disclose that we used AI tools for grammar and stylistic editing, structuring parts of the API specification, and for generating and refining request/response examples.
The authors carefully reviewed all AI-generated content.

\anon{We would like to thank Dr. Ho Shen Yong and Patrick Bassner for their support of our project.}
\end{acks}


\bibliographystyle{ACM-Reference-Format}
\bibliography{references}

@misc{solch2026uedapi,
 title={muEd-API}, url={osf.io/fet3u}, DOI={10.17605/OSF.IO/FET3U}, publisher={OSF}, author={Sölch, Maximilian and Neagu, Alexandra and Messer, Marcus and Johnson, Peter B and Kortemeyer, Gerd and Ng, Samuel S H and Lim, Fun S and Krusche, Stephan}, year={2026}, month={Feb} 
 }

@book{thomas2026critical,
  author    = {Thomas, Duncan A. and Laterza, Vito},
  title     = {Critical Perspectives on {EdTech} in Higher Education: Varieties of Platformisation},
  year      = {2026},
  publisher = {Springer Nature}
}

@article{Pangrazio02012023,
  author    = {Pangrazio, Luci and Selwyn, Neil and Cumbo, Bronwyn},
  title     = {A Patchwork of Platforms: Mapping Data Infrastructures in Schools},
  journal   = {Learning, Media and Technology},
  year      = {2023},
  publisher = {Routledge},
  volume    = {48},
  number    = {1},
  pages     = {65--80},
  doi       = {10.1080/17439884.2022.2035395}
}

@incollection{Dragoni2017,
  author    = {Dragoni, Nicola and Giallorenzo, Saverio and Lafuente, Alberto Lluch and Mazzara, Manuel and Montesi, Fabrizio and Mustafin, Ruslan and Safina, Larisa},
  title     = {Microservices: Yesterday, Today, and Tomorrow},
  booktitle = {Present and Ulterior Software Engineering},
  editor    = {Mazzara, Manuel and Meyer, Bertrand},
  year      = {2017},
  publisher = {Springer International Publishing},
  address   = {Cham},
  pages     = {195--216}
}

@article{batane2010turning,
  author    = {Batane, Tshepo},
  title     = {Turning to {Turnitin} to Fight Plagiarism Among University Students},
  journal   = {Journal of Educational Technology \& Society},
  year      = {2010},
  volume    = {13},
  number    = {2},
  pages     = {1--12},
  publisher = {JSTOR}
}

@inproceedings{Bakhouyi2017,
  author    = {Bakhouyi, Abdellah and Dehbi, Rachid and Talea, Mohamed and Hajoui, Omar},
  title     = {Evolution of Standardization and Interoperability on E-Learning Systems: An Overview},
  booktitle = {2017 16th International Conference on Information Technology Based Higher Education and Training (ITHET)},
  year      = {2017},
  pages     = {1--8},
  doi       = {10.1109/ITHET.2017.8067789}
}

@misc{lti,
  author       = {{1EdTech Consortium}},
  title        = {Learning Tools Interoperability ({LTI})},
  year         = {2026},
  howpublished = {Retrieved January 21, 2026 from \url{https://www.1EdTech.org/standards/lti}}
}

@misc{scorm,
  author       = {{Rustici Software}},
  year = {2026},
  title        = {{SCORM.com}: {SCORM} Explained},
  howpublished = {Retrieved January 21, 2026 from \url{https://scorm.com/}},
}

@misc{openapi,
  author       = {{OpenAPI Initiative}},
  title        = {{OpenAPI} Specification},
  year         = {2026},
  howpublished = {Retrieved January 21, 2026 from \url{https://www.openapis.org}}
}

@misc{xapi,
  author       = {{Rustici Software}},
  year = {2026},
  title        = {{xAPI.com}: Experience {API} ({xAPI}) Explained},
  howpublished = {Retrieved January 21, 2026 from \url{https://xapi.com/}},
}

@misc{openai_api,
  author       = {{OpenAI}},
  title        = {{OpenAI} {API}},
  year         = {2023},
  howpublished = {Retrieved January 21, 2026 from \url{https://github.com/openai/openai-openapi}}
}

@misc{mobius,
  author       = {{DigitalEd}},
  title        = {M\"{o}bius},
  year         = {2026},
  howpublished = {Retrieved April 12, 2026 from \url{https://www.digitaled.com/mobius/}}
}

@inproceedings{johnson2025lambdafeedback,
  author       = {Johnson, Peter B. and Fenton, Jon and Ramsden, Phil and Chatley, Robert and Ribera-Vicent, Maria and Lundeng{\aa}rd, Karl},
  title        = {Formative Feedback on Engineering Self-Study: Towards 1 Million Times per Year per Cohort},
  booktitle    = {2025 IEEE Global Engineering Education Conference (EDUCON)},
  year         = {2025},
  pages        = {1--3},
  organization = {IEEE}
}

@inproceedings{krusche2018artemis,
  author    = {Krusche, Stephan and Seitz, Andreas},
  title     = {Artemis: An Automatic Assessment Management System for Interactive Learning},
  booktitle = {Proceedings of the 49th ACM Technical Symposium on Computer Science Education},
  year      = {2018},
  publisher = {Association for Computing Machinery},
  address   = {New York, NY, USA},
  pages     = {284--289}
}

@article{lai2025NALA,
  author    = {Lai, Joel Weijia and Qiu, Wei and Thway, Maung and Zhang, Lei and Jamil, Nurabidah Binti and Su, Chit Lin and Ng, Samuel S. H. and Lim, Fun Siong},
  title     = {Leveraging Process-Action Epistemic Network Analysis to Illuminate Student Self-Regulated Learning with a Socratic Chatbot},
  journal   = {Journal of Learning Analytics},
  year      = {2025},
  volume    = {12},
  number    = {1},
  pages     = {32--49},
  publisher = {ERIC}
}

@article{kortemeyer2024ethel,
  author    = {Kortemeyer, Gerd},
  title     = {Ethel: A Virtual Teaching Assistant},
  journal   = {The Physics Teacher},
  year      = {2024},
  volume    = {62},
  number    = {8},
  pages     = {698--699},
  publisher = {AIP Publishing}
}

@inproceedings{Bloch2006,
  author    = {Bloch, Joshua},
  title     = {How to Design a Good {API} and Why It Matters},
  year      = {2006},
  publisher = {Association for Computing Machinery},
  address   = {New York, NY, USA},
  doi       = {10.1145/1176617.1176622},
  booktitle = {Companion to the 21st ACM SIGPLAN Symposium on Object-Oriented Programming Systems, Languages, and Applications},
  pages     = {506--507},
  location  = {Portland, Oregon, USA},
  series    = {OOPSLA '06}
}

@inproceedings{Stylos2007,
  author    = {Stylos, Jeffrey and Myers, Brad},
  title     = {Mapping the Space of {API} Design Decisions},
  booktitle = {IEEE Symposium on Visual Languages and Human-Centric Computing (VL/HCC 2007)},
  year      = {2007},
  pages     = {50--60},
  doi       = {10.1109/VLHCC.2007.44}
}

@misc{seoul,
  author       = {{Seoul Accord Secretariat}},
  title        = {The Seoul Accord},
  year         = {2026},
  howpublished = {Retrieved January 23, 2026 from \url{https://www.seoulaccord.org/}}
}

@article{lundengaard2024automated,
  author    = {Lundeng{\aa}rd, Karl and Johnson, Peter and Ramsden, Phil},
  title     = {Automated Feedback on Student Attempts to Produce a Set of Dimensionless Power Products from a Set of Physical Quantities That Describe a Physical Problem},
  journal   = {International Journal for Technology in Mathematics Education},
  year      = {2024},
  volume    = {31},
  number    = {3},
  pages     = {117--124},
  publisher = {Research Information}
}

@article{kerssens2022governed,
  author    = {Kerssens, Niels and Van Dijck, Jos{\'e}},
  title     = {Governed by Edtech? Valuing Pedagogical Autonomy in a Platform Society},
  journal   = {Harvard Educational Review},
  year      = {2022},
  volume    = {92},
  number    = {2},
  pages     = {284--303},
  publisher = {Harvard Education Publishing Group}
}

@misc{lewis2014definition,
  author       = {Lewis, James and Fowler, Martin},
  title        = {Microservices: A Definition of This New Architectural Term},
  year         = {2014},
  howpublished = {Retrieved January 21, 2026 from \url{https://martinfowler.com/articles/microservices.html}}
}

@article{Selwyn2023,
  author  = {Selwyn, Neil and Hillman, Thomas and Bergviken-Rensfeldt, Annika and Perrotta, Carlo},
  title   = {Making Sense of the Digital Automation of Education},
  journal = {Postdigital Science and Education},
  year    = {2023},
  volume  = {5},
  number  = {1},
  pages   = {1--14},
  doi     = {10.1007/s42438-022-00362-9}
}

@article{Zhang2024,
  author  = {Zhang, Xin and Zhang, Peng and Shen, Yuan and Liu, Min and Wang, Qiong and Ga{\v s}evi{\'c}, Dragan and Fan, Yizhou},
  title   = {A Systematic Literature Review of Empirical Research on Applying Generative Artificial Intelligence in Education},
  journal = {Frontiers of Digital Education},
  year    = {2024},
  volume  = {1},
  number  = {3},
  pages   = {223--245},
  doi     = {10.1007/s44366-024-0028-5}
}

@article{Holmes2022,
  author  = {Holmes, Wayne and Porayska-Pomsta, Kaska and Holstein, Ken and Sutherland, Emma and Baker, Toby and Buckingham Shum, Simon and Santos, Olga C. and Rodrigo, Mercedes T. and Cukurova, Mutlu and Bittencourt, Ig Ibert and Koedinger, Kenneth R.},
  title   = {Ethics of {AI} in Education: Towards a Community-Wide Framework},
  journal = {International Journal of Artificial Intelligence in Education},
  year    = {2022},
  volume  = {32},
  number  = {3},
  pages   = {504--526},
  doi     = {10.1007/s40593-021-00239-1}
}

@article{Klimova2023,
  author  = {Klimova, Blanka and Pikhart, Marcel and Kacetl, Jaroslav},
  title   = {Ethical Issues of the Use of {AI}-Driven Mobile Apps for Education},
  journal = {Frontiers in Public Health},
  year    = {2023},
  volume  = {10},
  doi     = {10.3389/fpubh.2022.1118116}
}

@article{FU2024100306,
  author  = {Fu, Yao and Weng, Zhenjie},
  title   = {Navigating the Ethical Terrain of {AI} in Education: A Systematic Review on Framing Responsible Human-Centered {AI} Practices},
  journal = {Computers and Education: Artificial Intelligence},
  year    = {2024},
  volume  = {7},
  pages   = {100306}
}

@article{Buckingham2023,
  author  = {Buckingham Shum, Simon and Lim, Lisa-Angelique and Boud, David and Bearman, Margaret and Dawson, Phillip},
  title   = {A Comparative Analysis of the Skilled Use of Automated Feedback Tools Through the Lens of Teacher Feedback Literacy},
  journal = {International Journal of Educational Technology in Higher Education},
  year    = {2023},
  volume  = {20},
  number  = {1},
  pages   = {40},
  doi     = {10.1186/s41239-023-00410-9}
}

@article{fullan2014rich,
  author    = {Fullan, Michael and Langworthy, Maria},
  title     = {A Rich Seam: How New Pedagogies Find Deep Learning},
  year      = {2014},
  publisher = {Pearson}
}

@book{reigeluth2013reinventing,
  author    = {Reigeluth, Charles M. and Karnopp, Jennifer R.},
  title     = {Reinventing Schools: It's Time to Break the Mold},
  year      = {2013},
  publisher = {Bloomsbury Publishing PLC}
}

@article{network2019scaling,
  author    = {{Omiyad Network}},
  title     = {Scaling Access \& Impact: Realizing the Power of {EdTech}. Executive Summary},
  year      = {2019},
  publisher = {Omiyad Network}
}

@article{Otto2022,
  author  = {Otto, Daniel and Kerres, Michael},
  title   = {Increasing Sustainability in Open Learning: Prospects of a Distributed Learning Ecosystem for Open Educational Resources},
  journal = {Frontiers in Education},
  year    = {2022},
  volume  = {7},
  doi     = {10.3389/feduc.2022.866917}
}

@techreport{Gottschalk2023,
  author      = {Gottschalk, Francesca and Weise, Crystal},
  title       = {Digital Equity and Inclusion in Education: An Overview of Practice and Policy in {OECD} Countries},
  institution = {OECD},
  year        = {2023},
  url         = {https://www.proquest.com/working-papers/digital-equity-inclusion-education-overview/docview/2849362537/se-2},
  note         = {Retrieved January 21, 2026}
}

@inproceedings{stocker2018interface,
  author    = {Stocker, Mirko and Zimmermann, Olaf and Zdun, Uwe and L{\"u}bke, Daniel and Pautasso, Cesare},
  title     = {Interface Quality Patterns: Communicating and Improving the Quality of Microservices {APIs}},
  booktitle = {Proceedings of the 23rd European Conference on Pattern Languages of Programs},
  year      = {2018},
  publisher = {Association for Computing Machinery},
  address   = {New York, NY, USA},
  pages     = {1--16}
}

@article{liu2016automated,
  author  = {Liu, Ming and Li, Yi and Xu, Weiwei and Liu, Li},
  title   = {Automated Essay Feedback Generation and Its Impact on Revision},
  journal = {IEEE Transactions on Learning Technologies},
  year    = {2016},
  volume  = {10},
  number  = {4},
  pages   = {502--513},
  publisher = {IEEE}
}

@inproceedings{VACALOPOULOU2024AI4,
  author    = {Vacalopoulou, A. and Gardelli, V. and Karafyllidis, T. and Liwicki, F. and Mokayed, H. and Papaevripidou, M. and Paraskevopoulos, G. and Stamouli, S. and Katsamanis, A. and Katsouros, V.},
  title     = {{AI4EDU}: An Innovative Conversational {AI} Assistant for Teaching and Learning},
  booktitle = {INTED2024 Proceedings},
  year      = {2024},
  publisher = {IATED},
  address   = {Valencia, Spain},
  pages     = {7119--7127},
  doi       = {10.21125/inted.2024.1877},
  series    = {18th International Technology, Education and Development Conference}
}

@inproceedings{solch:2025:DirectAutomatedFeedback,
  author    = {S\"{o}lch, Maximilian and Dietrich, Felix T. J. and Krusche, Stephan},
  title     = {Direct Automated Feedback Delivery for Student Submissions Based on {LLMs}},
  booktitle = {Proceedings of the 33rd ACM International Conference on the Foundations of Software Engineering},
  year      = {2025},
  month     = {Jul},
  publisher = {Association for Computing Machinery},
  address   = {New York, NY, USA},
  doi       = {10.1145/3696630.3727247},
  pages     = {901--911},
  series    = {FSE Companion '25}
}

@inproceedings{solch:2026:ScalingAssessmentStudent,
  author    = {S\"{o}lch, Maximilian and Krusche, Stephan},
  title     = {Scaling Assessment of Student Models with {LLMs}: Integrating Feedback into Practice},
  booktitle = {Proceedings of the 2026 IEEE/ACM 48th International Conference on Software Engineering},
  year      = {2026},
  publisher = {Association for Computing Machinery},
  address   = {New York, NY, USA},
  doi       = {10.1145/3786580.3786985},
  series    = {ICSE-SEET '26}
}

@article{granic2022educational,
  author    = {Grani{\'c}, Andrina},
  title     = {Educational Technology Adoption: A Systematic Review},
  journal   = {Education and Information Technologies},
  year      = {2022},
  volume    = {27},
  number    = {7},
  pages     = {9725--9744},
  publisher = {Springer}
}

@book{baldwin2000design,
  author    = {Baldwin, Carliss Y. and Clark, Kim B.},
  title     = {Design Rules, Volume 1: The Power of Modularity},
  year      = {2000},
  publisher = {MIT Press}
}

@misc{johnson2025microservices,
  author       = {Johnson, Peter and Ramsden, Phil and Messer, Marcus},
  title        = {{AI} Microservices for Sustainable Innovation in Education},
  year         = {2025},
  doi          = {10.35542/osf.io/wq4bd\_v3},
  note         = {EdArXiv preprint}
}

\end{document}